\documentclass[final,5p,times,twocolumn,nopreprintline]{elsarticle}
\usepackage{amsmath,slashed,booktabs,dsfont}
\usepackage{graphicx,graphics}
\usepackage{dcolumn}
\usepackage[hyperfootnotes=false,colorlinks=true]{hyperref}
\usepackage{xspace}
\usepackage{xcolor}
\usepackage{balance}
\usepackage{multirow}
\usepackage{multicol}

\usepackage{fancyhdr}
\fancypagestyle{firstpage}{%
	
	\lhead{}
	\rhead{\small}
}
\newcommand{\Order}{\mathcal{O}}
\newcommand{\Op}{\mathcal{O}}

\newcommand{\A}{\mathcal{A}}
\newcommand{\F}{\mathcal{F}}
\newcommand{\Hc}{\mathcal{H}}

\newcommand{\eps}{\epsilon}

\newcommand{\sth}{s_\text{th}}
\newcommand{\sps}{s_\text{ps}}
\newcommand{\muth}{\mu_\text{th}^2}

\renewcommand{\Im}{\text{Im}\,}
\newcommand{\disc}{\text{disc}\,}

\newcommand{\dilog}[1]{\mathrm{Li}_2\left(#1\right)}

\allowdisplaybreaks[1]

\begin{document}

\renewcommand{\theequation}{\arabic{equation}}

\begin{frontmatter}
 
\title{Reconciling hadronic and partonic analyticity in $b\to s\ell\ell$ transitions}

\author[Bern]{Martin Hoferichter}
\author[Bonn]{Bastian Kubis}
\author[Bonn]{Simon Mutke}

\address[Bern]{Albert Einstein Center for Fundamental Physics, Institute for Theoretical Physics, University of Bern, Sidlerstrasse 5, 3012 Bern, Switzerland}
\address[Bonn]{Helmholtz-Institut f\"ur Strahlen- und Kernphysik (Theorie) and
Bethe Center for Theoretical Physics, Universit\"at Bonn, 53115 Bonn, Germany}

\begin{abstract}
 Rare $B$-meson decays mediated by $b\to s\ell\ell$ transitions constitute sensitive probes of physics beyond the Standard Model, and have triggered considerable interest due to hints for deviations from the Standard-Model prediction. To establish a discrepancy beyond a reasonable doubt, control over the nonlocal matrix elements involving charm loops is essential, which, for large spacelike virtualities, can be constrained by an operator product expansion with coefficients known at two-loop order. We observe that the analytic structure of this partonic calculation, whose understanding is important to put forward rigorous parameterizations, follows from simple triangle topologies and demonstrate explicitly how dispersion relations are fulfilled even in the case of anomalous thresholds. Crucially, these anomalous contributions match onto the ones expected when considering hadronic degrees of freedom, proving that the partonic calculation does not miss anomalous effects and justifying its use in regions of parameter space in which a perturbative description applies.
\end{abstract}

\end{frontmatter}

\thispagestyle{firstpage}

\section{Introduction}

Observables mediated by $b\to s\ell\ell$ transitions are excellent probes of physics beyond the Standard Model (BSM), given that SM contributions are suppressed by loop factors and Cabibbo--Kobayashi--Maskawa matrix elements, and have triggered considerable interest due to various hints for discrepancies with the SM expectation.  Such hints persist for
 angular observables and decay rates in the $b\to s\mu\mu$ channel, displaying sizable deviations from their SM predictions~\cite{Descotes-Genon:2012isb,Descotes-Genon:2013vna,LHCb:2014cxe,LHCb:2020lmf,LHCb:2021zwz,Parrott:2022zte,Gubernari:2022hxn,CMS:2024atz,LHCb:2025mqb}. However, a frequent objection concerns the role of charm loops, whose matrix elements, encoded in so-called nonlocal form factors (FFs), need to be controlled to preclude that missed charm-loop effects mimic
 a BSM contribution~\cite{Beneke:2001at,Beneke:2004dp,Khodjamirian:2010vf,Khodjamirian:2012rm,Asatrian:2019kbk}.

For a given $B$-meson decay, the FFs of interest describe the $B\to (P,V)\,\gamma^*$ matrix element, with pseudoscalars $P=K,\pi,\ldots$, vectors $V=K^*,\phi,\rho,\omega,\ldots$, and the $\ell^+\ell^-$ pair attached to the virtual photon. These FFs are constrained by  analyticity and unitarity as well as various experimental inputs, such as the residues of the $J/\psi$ and $\psi(2S)$ poles, all of which can be incorporated in a dispersive approach~\cite{Bobeth:2017vxj,Gubernari:2020eft,Gubernari:2022hxn,Gubernari:2023puw,Gopal:2024mgb}. Indeed,  dispersive techniques have become increasingly relevant for the description of related hadronic matrix elements~\cite{Lyon:2014hpa,Cornella:2020aoq,Marshall:2023aje,Hanhart:2023fud,Bordone:2024hui}. Complementary to estimates using Lagrangian-based techniques~\cite{Isidori:2024lng,Isidori:2025dkp}, it is thus important to fully capture the analytic structure of the nonlocal FFs.

In this regard, a possible distortion of the analytic structure due to anomalous thresholds has become a concern~\cite{Ciuchini:2022wbq,Ladisa:2022vmh}, and the detailed analysis from Ref.~\cite{Mutke:2024tww}, using the example of the $u$-quark loop, demonstrates that indeed anomalous contributions do arise and can become sizable in certain circumstances. This analysis is based on a hadronic description via triangle topologies, in which case it is well understood how to set up a dispersive representation~\cite{Lucha:2006vc,Hoferichter:2013ama,Colangelo:2015ama,Hoferichter:2019nlq}. Extensions to the phenomenologically most relevant charm loop are in progress, to obtain a data-driven estimate of anomalous charm-loop effects in $B\to(P,V)\,\gamma^*$ FFs.

\begin{figure}[t]
	\centering
	\includegraphics[width=0.95\linewidth]{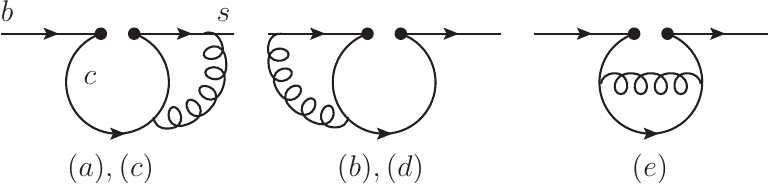}
	\caption{Basic partonic two-loop diagrams for $b\to s\gamma^*$ featuring charm loops, in the conventions of Ref.~\cite{Asatrian:2019kbk}. The straight lines indicate quarks, the black dots the insertion of the effective operator $\Op_{1,2}$, and the curly lines gluons, while the photon couples to these building blocks as follows: in diagrams $(a)$ and $(b)$, the electromagnetic current couples in all possible ways to the $b$ or $s$ quark; in diagrams $(c)$, $(d)$, and $(e)$ in all possible ways to the $c$ quark; type-$(e)$ diagrams with photon coupling at the $b$ or $s$ quark vanish.}
	\label{fig:diagrams_partonic}
\end{figure}

\begin{table*}[t]
\centering
\renewcommand{\arraystretch}{1.3}
\scalebox{0.95}{
\begin{tabular}{cccc}\toprule
Diagram $(a)$ & Diagram $(b)$ & Diagram $(c)$ & Diagram $(d)$\\
\includegraphics[width=4.3cm]{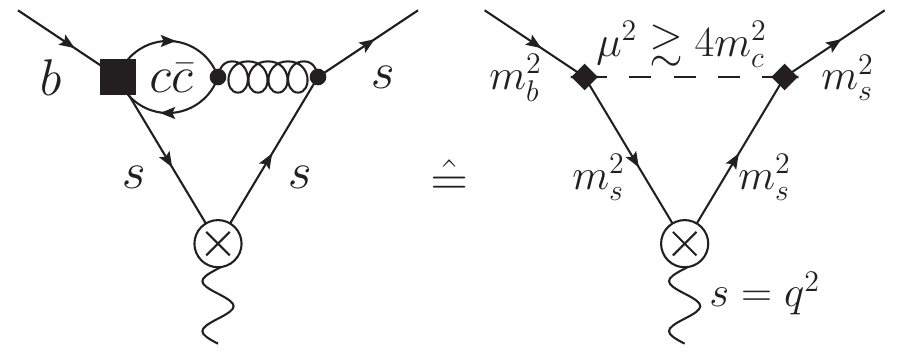}&
\includegraphics[width=4.3cm]{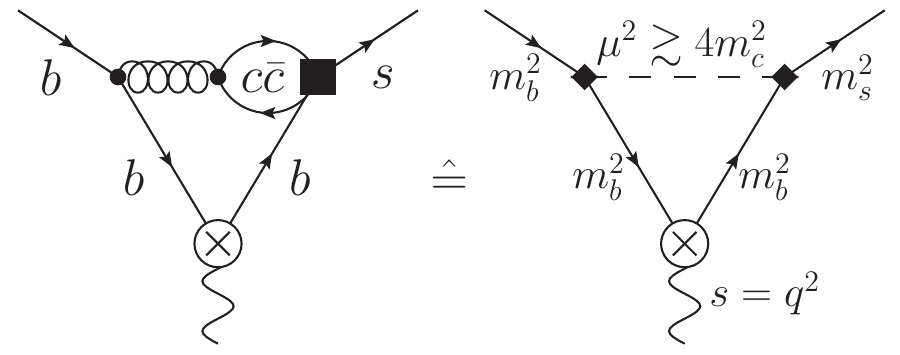}&
\includegraphics[width=4.3cm]{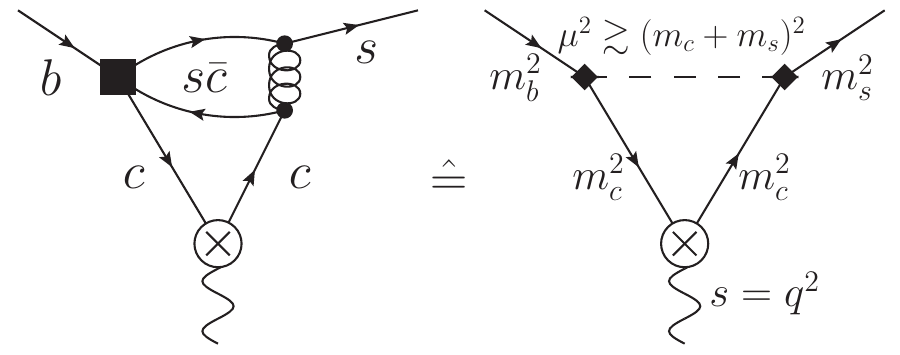}&
\includegraphics[width=4.3cm]{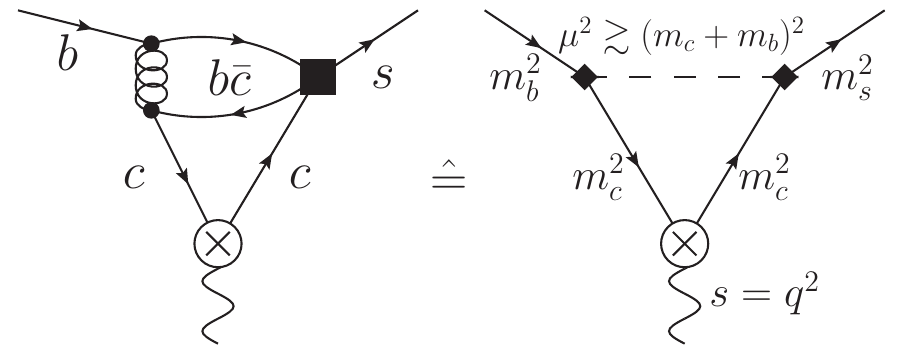}\\\midrule
$\sth=4m_s^2=0$ & $\sth=4m_b^2=4$ & $\sth=4m_c^2=0.4$ & $\sth=4m_c^2=0.4$\\
$\mu_\text{th}^2=4m_c^2=0.4$ & $\mu_\text{th}^2=4m_c^2=0.4$ & $\mu_\text{th}^2=(m_c+m_s)^2=0.1$&
$\mu_\text{th}^2=(m_c+m_b)^2\simeq 1.73$\\
$s_+=m_b^2-4m_c^2=0.6$ & \textcolor{gray}{$s_+=1.3+0.9i$}& $s_+=m_b^2=1$ & \textcolor{gray}{$s_+\simeq -0.24$}\\
$s_-=0$ & \textcolor{gray}{$s_-=1.3-0.9i$}& $s_-=m_b^2=1$ &\textcolor{gray}{$s_-\simeq -0.24$}\\
 \bottomrule
 \renewcommand{\arraystretch}{1.0}
\end{tabular}}
\caption{Diagrams $(a)$--$(d)$ in partonic form (left) and their interpretation in terms of triangle diagrams (right). In each case, we give the normal threshold $\sth$ and anomalous branch points $s_\pm$ in terms of the respective quark masses and evaluated for $m_c^2/m_b^2=0.1$, $m_b=1$ (and always $m_s=0$), following the conventions of Ref.~\cite{Asatrian:2019kbk}. The threshold of the spectral function is indicated by $\mu_\text{th}$. In each case, we only show the partonic diagram with the nontrivial analytic structure, omitting diagrams required to restore gauge invariance. Diagram $(e)$ does not lead to a triangle topology. The anomalous branch points for diagrams $(b)$ and $(d)$ (in gray) are provided for completeness, but do not play a role for the analytic structure on the first sheet.}
 \label{tab:diagrams}
\end{table*}

Meanwhile, another important constraint on the nonlocal contributions arises from the operator product expansion (OPE)~\cite{Grinstein:2004vb,Khodjamirian:2010vf,Beylich:2011aq,Bell:2014zya,Asatrian:2019kbk}, which allows for perturbative calculations in certain corners of parameter space. In particular, Ref.~\cite{Asatrian:2019kbk} presents an analytic two-loop evaluation including both $q^2$ and the charm mass $m_c$, based on which analyticity of the FFs for each set of separately gauge-invariant diagrams can be tested. While numerical results for all discontinuities were presented, a dispersion relation was only established for one of the simpler topologies, see diagram $(b)$ in Fig.~\ref{fig:diagrams_partonic}, in which case no singularities besides the unitarity cut at $q^2\geq 4m_c^2$ arise. In particular, the analytic structure of diagrams $(a)$ and $(c)$ was not fully understood, suspecting that in the latter case an anomalous branch point could play a role. In this situation, the question has been put forward if the partonic calculation actually includes the anomalous effects expected from the hadronic picture, i.e., if it fully captures the analytic structure of the nonlocal FFs. Evidently, an affirmative answer would be necessary to justify the use of partonic input even for large spacelike virtualities and be able to combine such constraints with dispersive parameterizations construed for hadronic degrees of freedom.

To address this question we proceed as follows: first, we condense each partonic diagram to a triangle topology, which allows us to put forward efficient parameterizations for all discontinuities. Second, we determine their coefficients by a fit to the numerical results from Ref.~\cite{Asatrian:2019kbk} (using the implementation in the \texttt{EOS} software~\cite{EOSAuthors:2021xpv} in version 1.0.16~\cite{danny_van_dyk_2025_15783309}), and demonstrate explicitly that all separately gauge-invariant classes of diagrams fulfill their own dispersion relation, even in those cases in which anomalous branch points play a role. Finally, having reduced the analytic structure of these diagrams to suitable triangle topologies, we show how the anomalous contributions in the partonic calculation map onto the ones in a hadronic picture. As key result, we can therefore reconcile the analyticity of nonlocal $b\to s\ell\ell$ FFs in hadronic and partonic descriptions, which justifies combined analyses that benefit both from data input in the timelike region and from perturbative constraints for large spacelike virtualities.

\begin{figure*}[t]
	\centering
	\includegraphics[width=0.35\linewidth]{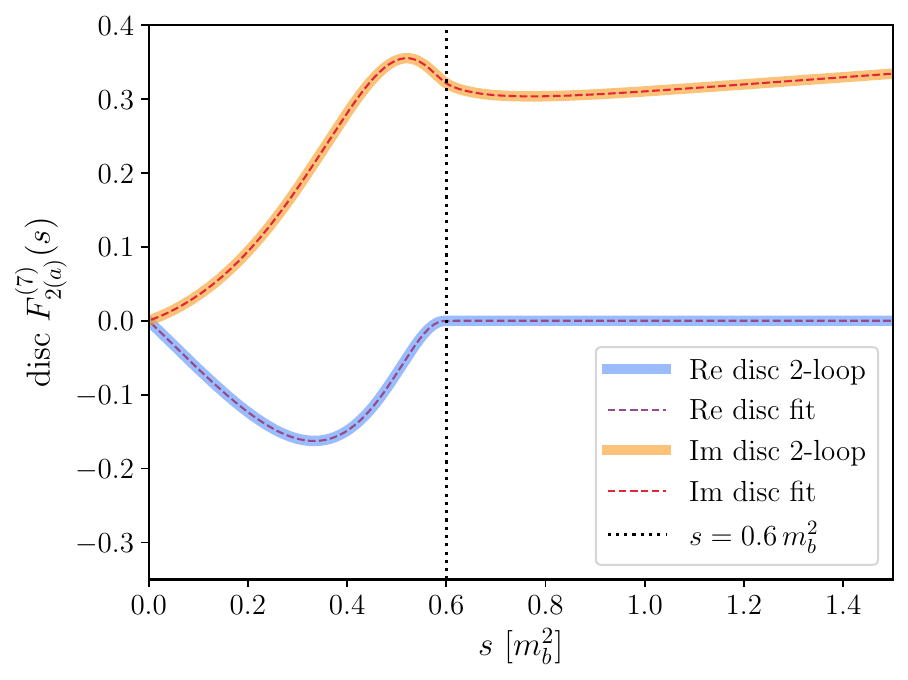}\quad
	\includegraphics[width=0.35\linewidth]{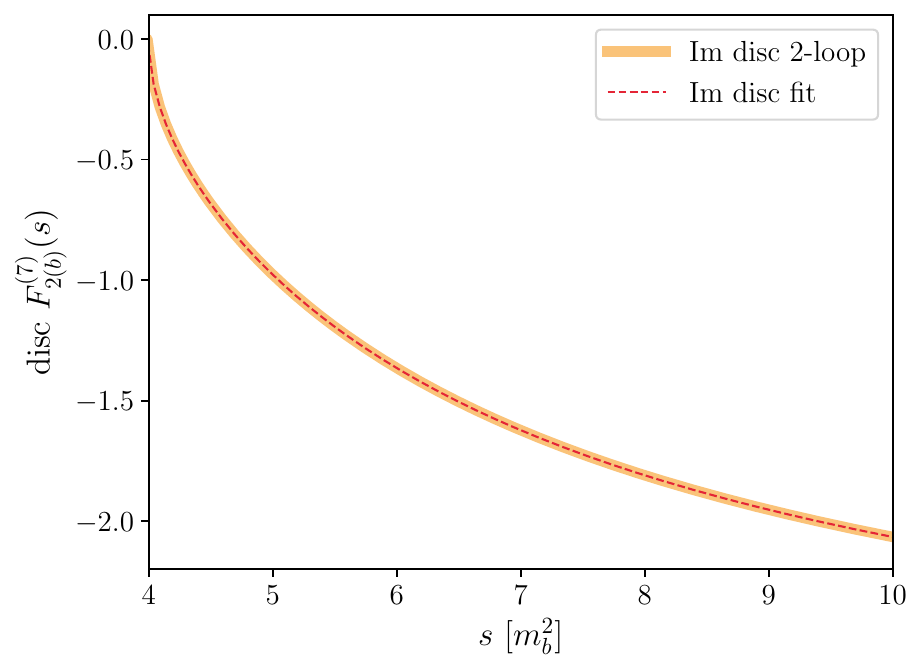}\\
	\includegraphics[width=0.35\linewidth]{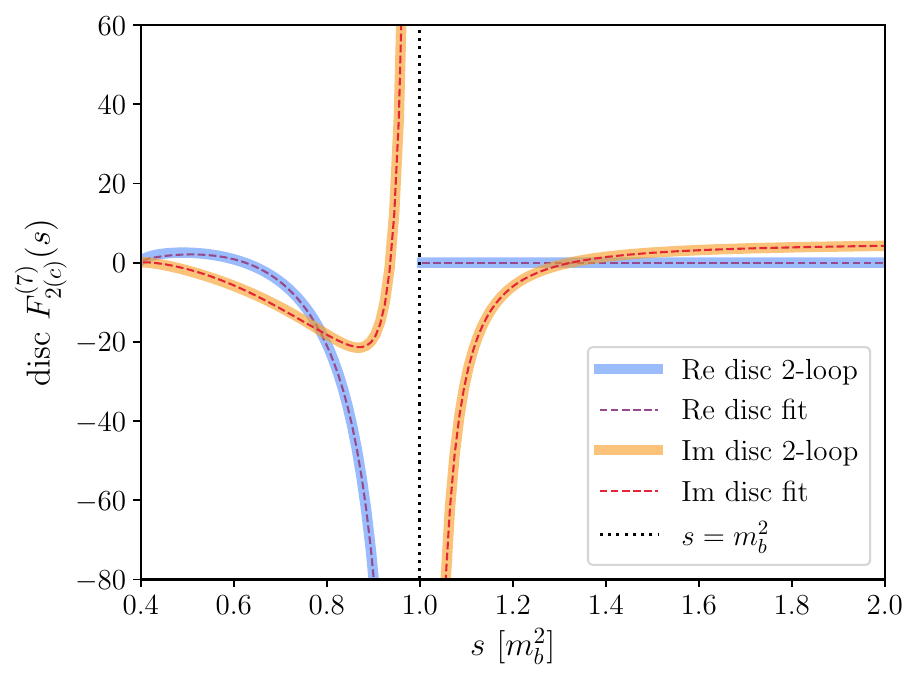}\quad
	\includegraphics[width=0.35\linewidth]{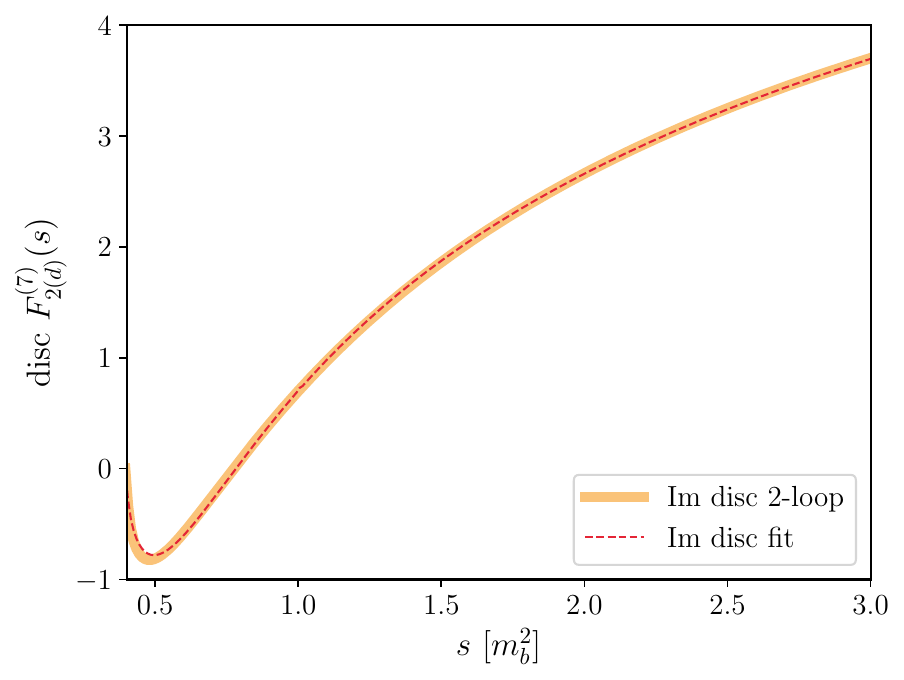}\\
	\caption{Discontinuities for diagrams $(a)$--$(d)$, comparing fits of our triangle-diagram-motivated parameterizations (``disc fit'') to the exact results from Ref.~\cite{Asatrian:2019kbk} (``disc $2$-loop''); see Ref.~\cite{EOS-DATA-2026-01}. For diagrams $(a)$ and $(c)$ the discontinuities develop a real part due to the anomalous thresholds, whose position is indicated by the vertical dotted lines. The plot ranges start at the normal threshold.}
	\label{fig:discontinuities}
\end{figure*}

\begin{figure*}[t]
	\centering
	\includegraphics[width=0.35\linewidth]{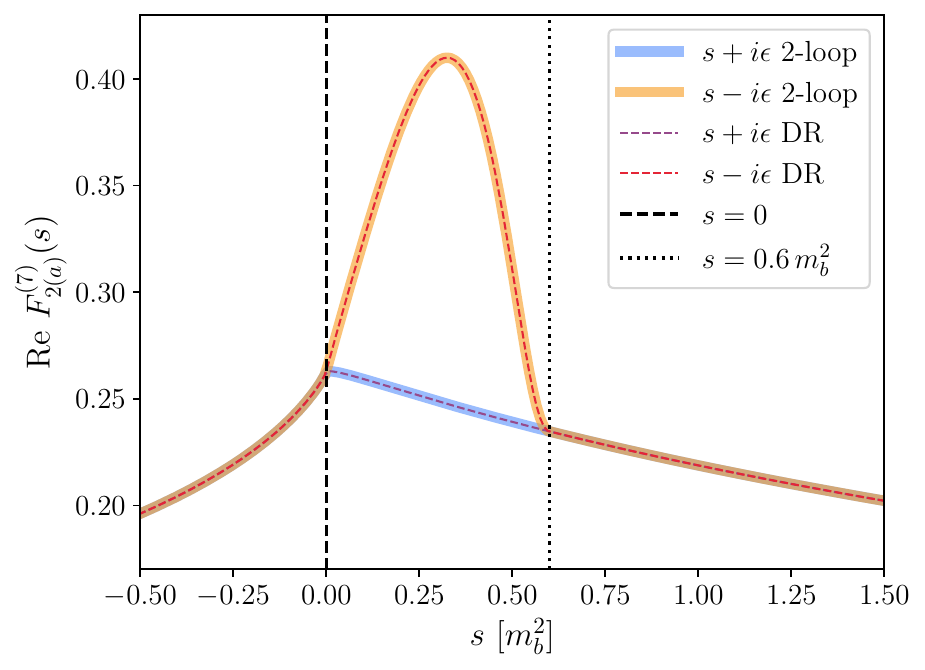}\quad
	\includegraphics[width=0.35\linewidth]{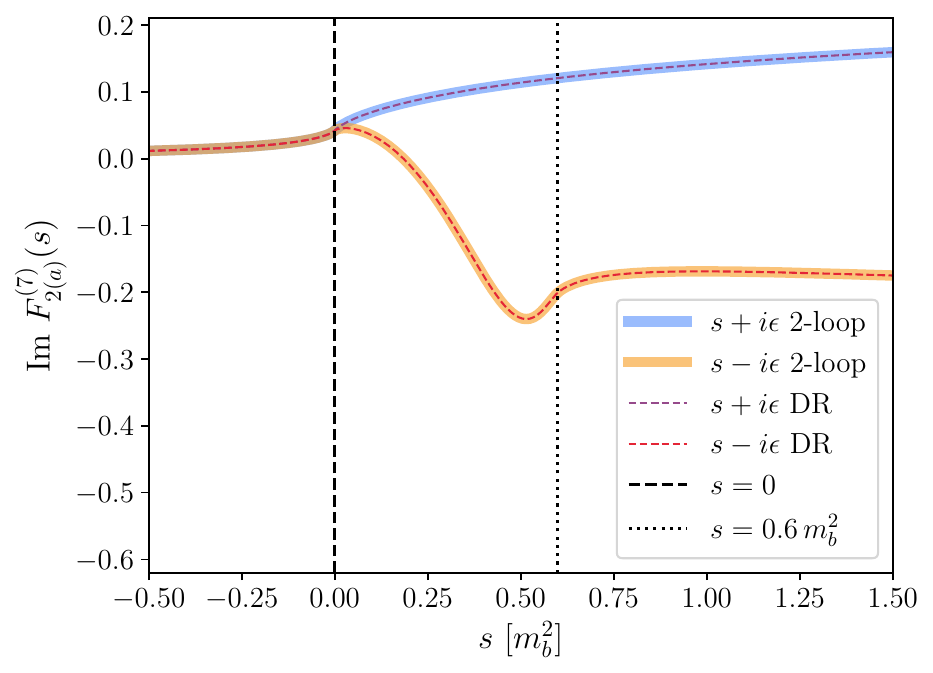}\\
	\includegraphics[width=0.35\linewidth]{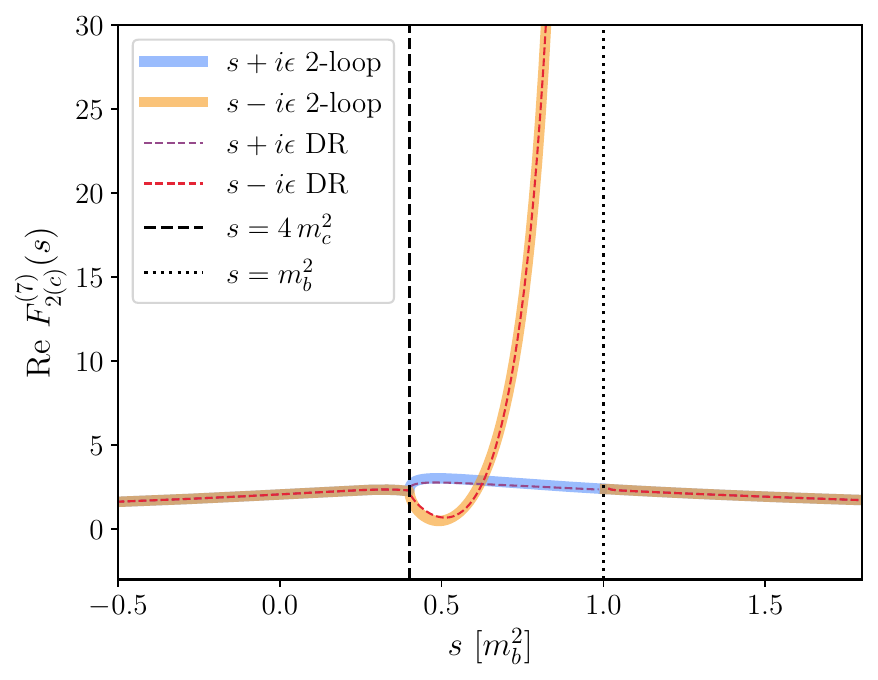}\quad
	\includegraphics[width=0.35\linewidth]{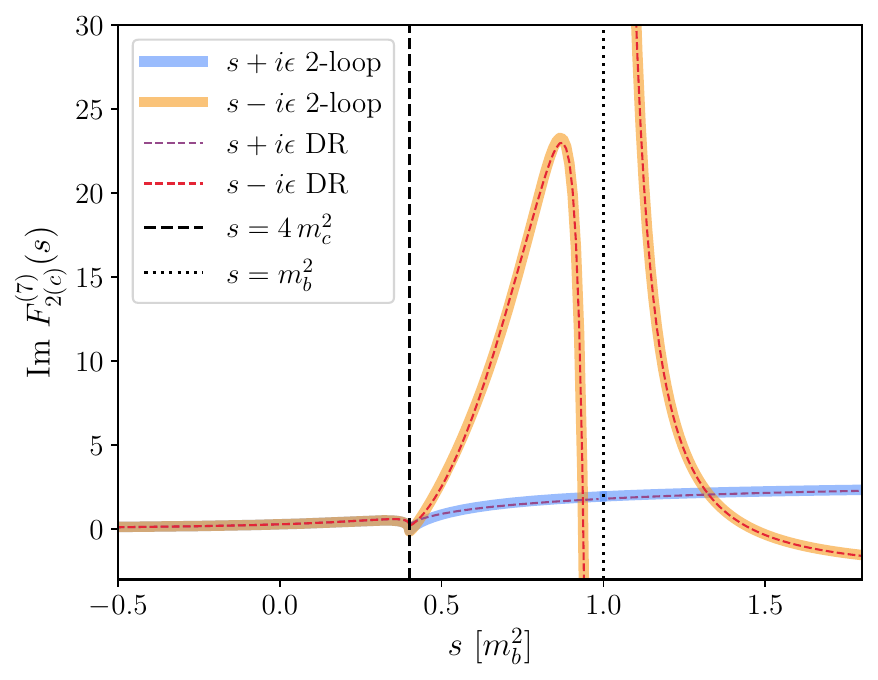}\\
	\caption{Real and imaginary part of the FFs for diagrams $(a)$ and $(c)$, comparing our results from the dispersion relation (``DR'') to the exact results from Ref.~\cite{Asatrian:2019kbk} (``$2$-loop''). The limits $s\pm i\eps$ disagree exactly where the respective discontinuity becomes nonzero; see Fig.~\ref{fig:discontinuities}. The results for diagrams $(b)$ and $(d)$ are shown in Fig.~\ref{fig:dispersion_relation_bd}. Vertical dashed and dotted lines indicate normal and anomalous thresholds, respectively. }
	\label{fig:dispersion_relation_ac}
\end{figure*}

\section{Formalism}

We decompose the $b\to s\ell\ell$ amplitude following the conventions of Ref.~\cite{Asatrian:2019kbk}, writing for the amplitude of a $B$-meson decaying into some meson $M$ and an $\ell^+\ell^-$ pair
\begin{align}
\A(\bar B\to M\ell^+\ell^-)
&=\frac{\alpha G_F V_{ts}^*V_{tb}}{\sqrt{2}\,\pi}\bigg[\big(C_9 L_V^\mu+C_{10}L_A^\mu\big)\F_\mu
\notag\\
&-\frac{L_V^\mu}{q^2}\Big\{2im_b C_7\F_\mu^T+\Hc_\mu\Big\}
\bigg],
\end{align}
where $q^2 = (q_1+q_2)^2 \equiv s$ is the invariant mass squared of the lepton pair, $L_{V(A)}^\mu=\bar u_\ell(q_1)\gamma^\mu(\gamma_5)v_\ell(q_2)$, $C_{7,9,10}$ are (effective) Wilson coefficients, and the local ($\F_\mu$, $\F_\mu^T$) and nonlocal ($\Hc_\mu$) FFs are defined by the matrix elements
\begin{align}
 \F_\mu&=\big\langle M(k)\big|\bar s\gamma_\mu P_L b\big|\bar B(q+k)\big\rangle,\\
 \F_\mu^T&=\big\langle M(k)\big|\bar s\sigma_{\mu\nu}q^\nu P_R\big|\bar B(q+k)\big\rangle,\notag\\
 \Hc^\mu&=16\pi^2 i\int d^4x\,e^{iq\cdot x}\notag\\
 &\times\big\langle M(k)\big|T\big\{j_\text{em}^\mu(x),\big(C_1\Op_1+C_2\Op_2\big)(0)\big\}\big|\bar B(q+k)\big\rangle,\notag
\end{align}
where $j_\text{em}^\mu$ is the electromagnetic current and $\Op_{1,2}$ refer to four-fermion operators  of flavor content $b\bar{s}c\bar{c}$. At low hadronic recoil,  the OPE relates the nonlocal to the local FFs via
\begin{align}
 \Hc_\mu^\text{OPE}&=\Delta C_9\big(q^2\big)\big(q_\mu q_\nu-q^2 g_{\mu\nu}\big)\F^\nu+2im_b\Delta C_7\big(q^2\big)\F^T_\mu,
\end{align}
up to subleading corrections, and the $q^2$-dependent shifts in the Wilson coefficient $\Delta C_{7,9}(q^2)$ are calculable in perturbation theory. In this work, we are interested in the analytic structure of the next-to-leading-order corrections computed in Ref.~\cite{Asatrian:2019kbk}.  Taking out the Wilson coefficients $C_{1,2}$ and a factor $-\alpha_s/(4\pi)$, this leads one to consider FFs $F_{i,(k)}^{(j)}(s)$, with $i\in\{1,2\}$, $j\in\{7,9\}$, and $k\in\{a,b,c,d,e\}$ referring to the diagrams shown in Fig.~\ref{fig:diagrams_partonic}. These five classes are separately electromagnetically gauge invariant (as can be shown by studying their dependence on the electric charges of the different quark flavors), and can therefore be studied on their own. Since the results for $i=1,2$ are related by simple color factors, while the analytic structure for $j=7,9$ is very similar, we follow Ref.~\cite{Asatrian:2019kbk} and concentrate on $F_{2,(k)}^{(7)}(s)$.

We posit that the analytic structure of each class can be understood in terms of a suitably chosen one-loop triangle diagram,  identified as in Table~\ref{tab:diagrams}, so that the form of the respective discontinuities
\begin{equation}
 \disc F_{2,(k)}^{(7)}(s)
 \equiv F_{2,(k)}^{(7)}(s+i\eps)-
 F_{2,(k)}^{(7)}(s-i\eps)
\end{equation}
derives from the general analysis in Ref.~\cite{Mutke:2024tww}. The simplest case is given by diagram $(b)$, for which only the normal threshold $\sth$ plays a role, with a discontinuity that behaves as $\sqrt{s-\sth}$. In this case, a dispersion relation was already established in Ref.~\cite{Asatrian:2019kbk} based on an empirical parameterization of the discontinuity.\footnote{This parameterization does not impose $\disc F_{2,(b)}^{(7)}(\sth)=0$, so that the resulting dispersion relation would diverge for $s\to\sth$.}  Here, we construct the discontinuities based on the expressions from Ref.~\cite{Mutke:2024tww}, introducing a spectral function $\rho(\mu^2)$ that describes the structure of the left-hand cut in terms of a variable $\mu^2$, whose threshold value $\mu_\text{th}^2$ again follows from an analysis of the respective partonic diagram; see Table~\ref{tab:diagrams}.
Diagram $(d)$ can be treated similarly, the main difference concerning a threshold divergence of $\disc F_{2,(d)}^{(7)}(s)$ that arises from the crossed-channel gluon exchange~\cite{Yennie:1961ad,Gasser:2007zt,Bissegger:2008ff}. The details of the construction of the discontinuity in both cases are described in \ref{app:discontinuities}.

Diagrams $(a)$ and $(c)$ are qualitatively different, with additional features already observed in Ref.~\cite{Asatrian:2019kbk} leading to discontinuities that are no longer purely imaginary for $s\in[0,0.6]$ and $s\in[0.4,1]$, respectively, and even exhibit a pole at $s=1$ in the case of $\disc F_{2,(c)}^{(7)}(s)$.\footnote{For better comparison with Ref.~\cite{Asatrian:2019kbk}, we use the same numerical values $m_c^2/m_b^2=0.1$, $m_b=1$.} These features can be explained precisely by the anomalous thresholds $s_\pm$ of the triangle diagrams assigned in Table~\ref{tab:diagrams}: for $\disc F_{2,(a)}^{(7)}(s)$, $s_+=0.6$ is located right on the unitarity cut, so that approaching the real axis from below between $s_+$ and $s_-=0$, an anomalous contribution is generated. Similarly, an anomalous contribution arises for $\disc F_{2,(c)}^{(7)}(s)$ between $s_+=1$ and $\sth=0.4$, with the additional complication that $s_+$ now coincides with the pseudothreshold $s_\text{ps} = (m_b-m_s)^2 = 1$; see \ref{app:discontinuities} and Ref.~\cite{Mutke:2024tww} for more details.

With discontinuities parameterized accordingly, in particular spectral functions $\rho(\mu^2)$ expanded into conformal polynomials, we obtain the fits shown in Fig.~\ref{fig:discontinuities}, in comparison to the exact results. Throughout, we observe that the discontinuities are well reproduced using our parameterizations derived from the triangle diagrams in Table~\ref{tab:diagrams}. In particular, the results for diagrams $(a)$ and $(c)$ reproduce the real parts generated due to the anomalous thresholds $s_+$, as well as the singularity structure around $s_+$ in the case of diagram $(c)$. The latter is related to the crossed-channel gluon exchange~\cite{Yennie:1961ad,Gasser:2007zt,Bissegger:2008ff}, and can be taken into account by means of the spectral function constructed in \ref{app:discontinuities}.

\section{Partonic dispersion relations}

With the discontinuities determined accordingly, analyticity of the FFs $F_{2,(k)}^{(7)}(s)$ demands that the full results can be reconstructed by a dispersion relation
\begin{equation}
\label{disp_relation}
F_{2,(k)}^{(7)}(s)=
F_{2,(k)}^{(7)}(s_0)
+\frac{s-s_0}{2\pi i}\int_{\sth}^\infty ds'\frac{\disc F_{2,(k)}^{(7)}(s')}{(s'-s_0)(s'-s)},
\end{equation}
where a subtraction at $s_0$ becomes necessary because, asymptotically, the discontinuities approach a constant. In practice, we set $s_0=-m_c^2$ for diagram $(a)$ to avoid the unitarity cut starting at $\sth=0$, while for all other diagrams we use $s_0=0$ for simplicity. For diagrams $(a)$, $(b)$, and $(d)$ the evaluation of the dispersion relation is straightforward, while for diagram $(c)$ more care is required. In this case, the anomalous threshold at $s_+=m_b^2$ coincides with the pseudothreshold, which complicates the singularity structure and leads to the more involved integration strategy described in \ref{app:dispersion_relations}.

The results of the dispersion relation for $F_{2,(a,c)}^{(7)}(s)$ are compared to the exact results in Fig.~\ref{fig:dispersion_relation_ac}, demonstrating that the partonic FFs indeed fulfill the expected dispersion relation once the anomalous contributions are properly taken into account.
Note that due to the intricate singularity structure in diagram $(c)$, it is not possible to separate normal and anomalous contributions to the dispersion relation in a regulator-independent manner, as only the sum yields a well-defined result, and already for diagram $(a)$ the normal and anomalous contributions display a logarithmic singularity at $s_+$, which can lead to $\Order(1)$ corrections~\cite{Mutke:2024tww}.
The corresponding results for diagrams $(b)$ and $(d)$, as well as comparisons in the complex $s$ plane, are provided in \ref{app:dispersion_relations_results}.

\section{Comparison to hadronic picture}

\begin{figure}[t]
	\centering
	\includegraphics[width=\linewidth]{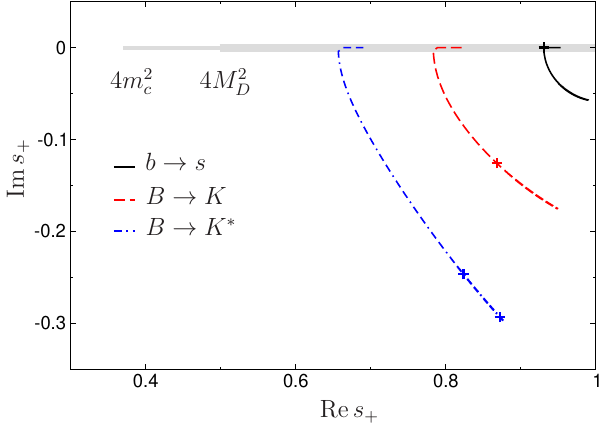}
	\caption{Trajectory of $s_+$ in the complex plane as a function of the parameter $\xi\in[0.8,10]$; see main text. The partonic threshold $\sth=4m_c^2$ and trajectory ($b\to s$, black) are given in units of $m_b$, the hadronic threshold $\sth=4M_D^2$ and trajectories ($B\to K$, red; $B\to K^*$, blue) in units of $M_B$. For $\xi=1$, $s_+$ moves onto the real axis. The crosses indicate the physical points, which occur at $\xi=1$ ($b\to s$, $m_1=m_c+m_s$), $\xi=2.04$ ($B\to K$, $m_1=M_{D^*_s}$), and $\xi=\{3.69,8.62\}$ ($B\to K^*$, $m_1=\{M_{D^*_s},M_{D_s}\}$), respectively. All masses are taken from Ref.~\cite{ParticleDataGroup:2024cfk}.}
	\label{fig:sp_trajectory}
\end{figure}

For a phenomenological description of the charm-loop contributions, the configuration in diagram $(c)$ becomes most relevant, as this topology maps onto a dispersion relation that relies on $\bar D D$ intermediate states. In particular, the question arises how the anomalous threshold at $s_+=m_b^2$ in the partonic picture maps onto the ones in the hadronic description, which were shown to lie in the lower complex plane in Ref.~\cite{Mutke:2024tww}. However, these anomalous thresholds in the partonic and hadronic description are, in fact, in a direct correspondence, and the only difference concerns hadronization and strange-quark mass effects. Restoring the latter, the partonic anomalous threshold changes to $s_+=m_c(m_b^2-m_s^2)/(m_c+m_s)<m_b^2$ at $\muth$, and for $\mu^2>\muth$ its position remains on the real axis. If, instead, values of $\mu^2<\muth$ were allowed, $s_+$ would become complex, and the corresponding trajectory is most conveniently parameterized in terms of
\begin{equation}
 \xi=\frac{\sqrt{p_3^2}}{m_1-m_3}=\frac{m_s}{\mu-m_c},
\end{equation}
using the general labeling from Fig.~\ref{fig:triangle}. If $\xi\leq 1$, the particle that describes the left-hand cut is kinematically allowed to decay, leading to a real value of $s_+$, otherwise, $s_+$ becomes complex. The resulting trajectories shown in Fig.~\ref{fig:sp_trajectory} for the partonic case $b\to s$ as well as $B\to\{K,K^*\}$ are indeed very similar, and mainly differ by the physical values of $m_s^2 < M_K^2 < M_{K^*}^2$. The location of $s_+$ only appears different because in the partonic case $\xi\leq 1$, while in the hadronic case the fact that $M_K$ and $M_{K^*}$ are larger than the mass difference between $D_{s}^{(*)}$ and $D$ mesons leads to values $\xi>1$.
This discussion shows that the anomalous contribution in the dispersion relation for diagram $(c)$ matches onto the anomalous effects identified in the hadronic picture in Ref.~\cite{Mutke:2024tww}, with differences in the trajectories of $s_+$ in the complex plane, see Fig.~\ref{fig:sp_trajectory}, solely driven by the phenomenology of the strange quark.

\section{Conclusions}

In this work we studied the analytic structure of the FFs that determine the nonlocal contributions to $b\to s\ell\ell$ in the OPE limit, calculated at two-loop order in Ref.~\cite{Asatrian:2019kbk}. While the general analysis of these two-loop diagrams, reproduced in Fig.~\ref{fig:diagrams_partonic}, is complicated, we argued that their main features can be described by mapping each topology onto one-loop triangle diagrams according to Table~\ref{tab:diagrams}. We found that the analytic structure is indeed reproduced exactly, crucially, once anomalous thresholds in diagrams $(a)$ and $(c)$, whose properties follow directly from the analysis in terms of triangle diagrams, are taken into account. In particular, we put forward a parameterization of the discontinuities derived from the respective triangle diagrams and demonstrated explicitly that the dispersion relation for each diagram is fulfilled, again, once the anomalous contributions are properly taken into account. Finally, we observed that the anomalous thresholds encountered in the partonic picture are in a one-to-one correspondence to the hadronic anomalous thresholds discussed in Ref.~\cite{Mutke:2024tww}, with differences driven by strange-quark mass effects and its hadronization into $K^{(*)}$ and $D_s^{(*)}$ mesons. Our results therefore demonstrate that the partonic calculation does not miss anomalous effects, as sometimes suggested in the discussion of nonlocal matrix elements in $b\to s\ell\ell$ transitions, justifying usage of the OPE constraint in those regions of parameter space in which the perturbative expansion applies.

\section*{Acknowledgments}

We thank C.~Greub, N.~Gubernari, M.~Reboud, D.~van Dyk, and J.~Virto for valuable discussions and comments on the manuscript. We further thank J.~Virto for help with the code of Ref.~\cite{Asatrian:2019kbk}, and M.~Reboud and D.~van Dyk with its implementation in the \texttt{EOS} software~\cite{EOSAuthors:2021xpv}, version 1.0.16~\cite{danny_van_dyk_2025_15783309}.
Financial support by the Swiss National Science Foundation (Project No.\  TMCG-2\_213690),
by the German Academic Scholarship Foundation,
and by the MKW NRW under funding code NW21-024-A
is gratefully acknowledged.

\appendix

\section{Discontinuities}
\label{app:discontinuities}
\setcounter{figure}{0}

\begin{figure}[t]
	\centering
	\includegraphics[width=0.4\linewidth]{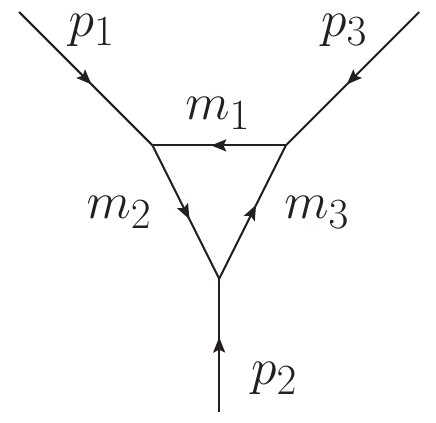}
	\caption{Conventions for the triangle diagram.}
	\label{fig:triangle}
\end{figure}

Starting point for the parameterization of the discontinuities is the triangle diagram shown in Fig.~\ref{fig:triangle}.
Considering first diagram $(a)$, the easiest functional form for the discontinuity, requiring the correct behavior at all thresholds, reads
\begin{align}
	\disc F_{2,(a)}^{(7)}\big(s;\mu^2\big)&=\disc F\big(s;\mu^2\big) \equiv 2i\frac{s^2}{\big[\kappa(s)\big]^2} \bigg( Y\big(s;\mu^2\big) \notag\\
	&+ \frac{\big[Y\big(s;\mu^2\big)\big]^2 - \big[\kappa(s)\big]^2}{2\kappa(s)} L\big(s;\mu^2\big) \bigg),
\end{align}
in terms of the general expressions
\begin{align}
	\kappa(s) &= \lambda^{1/2}\big(s,m_2^2,m_3^2\big) \, \lambda^{1/2}\big(s,p_1^2,p_3^2\big), \notag\\
	Y\big(s;\mu^2\big) &= s^2 - s \big(p_1^2+p_3^2+m_2^2+m_3^2-2\mu^2\big) \notag\\
	&+ \big(p_1^2-p_3^2\big)\big(m_2^2-m_3^2\big), \notag\\
	s_\pm\big(\mu^2\big)&=p_1^2\frac{\mu^2+m_3^2}{2\mu^2}+p_3^2\frac{\mu^2+m_2^2}{2\mu^2}-\frac{p_1^2p_3^2}{2\mu^2}\notag\\
	&-\frac{\big(\mu^2-m_2^2\big)\big(\mu^2-m_3^2\big)}{2\mu^2}\notag\\
	&\pm
	\frac{1}{2\mu^2}\lambda^{1/2}\big(p_1^2,\mu^2,m_2^2\big) \, \lambda^{1/2}\big(p_3^2,\mu^2,m_3^2\big),
\end{align}
evaluated for $m_2=m_3=m_s=0$, $p_1^2=m_b^2$, $p_3^2=m_s^2=0$, and the logarithm analytically continued as
\begin{equation}
\label{logarithm_diagram_a}
L\big(s;\mu^2\big) = \log \ \frac{Y\big(s;\mu^2\big)+\kappa(s)}{Y\big(s;\mu^2\big)-\kappa(s)} - i\pi  \, \theta\big[s_+\big(\mu^2\big)-s\big].
\end{equation}
In particular, $s_+(\mu^2)=m_b^2-\mu^2$, $s_-=0$, as given in Table~\ref{tab:diagrams}. Finally, the $\mu^2$ dependence is parameterized in terms of a spectral function via
\begin{equation}
 \disc F_{2,(a)}^{(7)}(s)=\int_{\muth}^\infty d\mu^2 \rho\big(\mu^2\big) \disc F_{2,(a)}^{(7)}\big(s;\mu^2\big),
\end{equation}
where $\muth=4m_c^2$ and
\begin{align} \label{F2a7_spectralfunction}
	\rho\big(\mu^2\big) &= \Im\bigg[\Big(z\big(\mu^2\big)-1\Big) \sum_{k=0}^{3} a_k \, \big[z\big(\mu^2\big)\big]^k \bigg],\notag\\
z\big(\mu^2\big) &\equiv z\big(\mu^2,\muth,\mu_0^2\big) = \frac{\sqrt{\muth-\mu^2}-\sqrt{\muth-\mu_0^2}}{\sqrt{\muth-\mu^2}+\sqrt{\muth-\mu_0^2}}.
\end{align}
In practice, we set $\mu_0^2=0$ and fit the coefficients $a_k$ to the  results of Ref.~\cite{Asatrian:2019kbk}.

For diagram $(b)$ we have
\begin{equation}
	\disc F_{2,(b)}^{(7)}\big(s;\mu^2\big)= \sigma_b(s) \disc F\big(s;\mu^2\big),\qquad \sigma_b(s)=\sqrt{1-\frac{4m_b^2}{s}},
\end{equation}
where now $m_2=m_3=m_b$, the complex logarithm takes the same form as in Eq.~\eqref{logarithm_diagram_a} without the Heaviside function,  and $s_\pm(\mu^2)$ are located on the second Riemann sheet. We use the same spectral function as in Eq.~\eqref{F2a7_spectralfunction}, apart from $\mu_0^2=-4m_c^2$.

For diagram $(c)$ the discontinuity reads
\begin{equation}
\label{disc_c}
	\disc F_{2,(c)}^{(7)}\big(s;\mu^2\big)= \sigma_c(s) \disc F\big(s;\mu^2\big),\qquad \sigma_c(s)=\sqrt{1-\frac{4m_c^2}{s}},
\end{equation}
where now $m_2=m_3=m_c$ and the logarithm needs to be analytically continued via
\begin{equation}
    L\big(s;\mu^2\big) = \log  \frac{Y\big(s;\mu^2\big)+\kappa(s)}{Y\big(s;\mu^2\big)-\kappa(s)} - i\pi \, \theta\big[s_+\big(\mu^2\big)-s\big]\,\theta\big[s-s_-\big(\mu^2\big)\big].
\end{equation}
The resulting discontinuity is finite for all $\mu^2>\muth=m_c^2$, but for $\mu^2=\muth$ one obtains a pole at $s=m_b^2$
\begin{equation}
	\disc F_{2,(c)}^{(7)}\big(s;\muth\big) = \frac{2i}{\sigma_c(s)\,(s-m_b^2)} \bigg[ s + \frac{2m_c^2}{\sigma_c(s)} \log\frac{1-\sigma_c(s)}{1+\sigma_c(s)} \bigg].
\end{equation}
This pole at $\mu^2=\muth$ needs to be reflected by the parameterization of the spectral function, which we decompose into a regular and divergent part according to
\begin{align}
\rho\big(\mu^2\big)&=\rho_\text{reg}\big(\mu^2\big)+\rho_\text{div}\big(\mu^2\big) \notag\\
&=\Im\bigg[\Big(z\big(\mu^2\big)-1\Big)^2 \sum_{k=0}^{6} a_k \, \big[z\big(\mu^2\big)\big]^k \bigg] +b_0 \frac{\log\Bigl(1-\frac{\muth}{\mu^2}\Bigr)}{\mu^2-\muth},
\end{align}
where we use $\mu_0^2=-m_c^2$ and where the integration over $\rho_\text{div}(\mu^2)$, whose form is motivated by the divergence structure of virtual Coulombic exchanges~\cite{Yennie:1961ad,Gasser:2007zt,Bissegger:2008ff}, diverges. We regulate this divergence by defining
\begin{align}
 \disc F_{2,(c)}^{(7)}(s)&=\int_{\muth}^\infty d\mu^2\rho_\text{reg}\big(\mu^2\big)\disc F_{2,(c)}^{(7)}\big(s;\mu^2\big)\notag\\
 &+\int_{\muth}^\infty d\mu^2\rho_\text{div}\big(\mu^2\big)G\big(s;\mu^2\big),
\end{align}
where
\begin{equation}
 G\big(s;\mu^2\big) = \frac{(s-m_b^2)+i(\mu^2-\muth)}{(s-m_b^2)^2 + \big(\mu^2-\muth\big)^2} \,\big(s-m_b^2\big)\disc F_{2,(c)}^{(7)}\big(s;\muth\big).
\end{equation}
Due to
\begin{equation}
\label{G_spectral_integral}
    \int_{\muth}^{\infty} d\mu^2 \frac{G(s;\mu^2)}{\mu^2-\muth} \simeq \Bigg[c_1 \log\bigg(\frac{s}{m_b^2}-1\bigg) + c_2\Bigg] \disc F_{2,(c)}^{(7)}\big(s;\muth\big),
\end{equation}
this form indeed matches the singularity structure observed in Ref.~\cite{Asatrian:2019kbk}. The dispersion integral over this singular spectral function remains well defined, and can be evaluated with the techniques described in \ref{app:dispersion_relations}.

For diagram $(d)$, the discontinuity takes a similar form as Eq.~\eqref{disc_c}, without the Heaviside function in the definition of the logarithm and where $s^2 \mapsto m_b^2 s$ in $\disc F(s;\mu^2)$. The spectral function is parameterized as
\begin{align}
    \rho\big(\mu^2\big) &= \Im\bigg[\Big(z\big(\mu^2\big)-1\Big)^2 \sum_{k=0}^{4} a_k \, \big[z\big(\mu^2\big)\big]^k \bigg]\notag\\
    &+b_0 \frac{\log\Bigl(\frac{\sqrt{(\mu^2-\muth)(\mu^2-\mu_-^2)}}{\mu^2}\Bigr)}{\sqrt{\bigl(\mu^2-\muth\bigr)\bigl(\mu^2-\mu_-^2\bigr)}},
\end{align}
where $\muth = (m_b + m_c)^2$, $\mu_-^2 = (m_b - m_c)^2$, and $\mu_0^2=0$.

\section{Implementation of the dispersion relations}
\label{app:dispersion_relations}
\setcounter{figure}{0}

For diagrams $(a)$ and $(b)$, the evaluation of the dispersion relation~\eqref{disp_relation} is straightforward once the parameterization of the discontinuity is fit to the results from Ref.~\cite{Asatrian:2019kbk} (using the implementation  in the \texttt{EOS} software~\cite{EOSAuthors:2021xpv} in version 1.0.16~\cite{danny_van_dyk_2025_15783309}). In the case of diagram $(d)$, the integration over the spectral function involves a singularity at $\mu^2=\muth$, but the threshold singularity $\propto 1/\sqrt{\mu^2-\muth}$ is integrable.
In contrast, for diagram $(c)$ the fact that the anomalous threshold at $s_+=m_b^2$ coincides with the pseudothreshold $\sps$ requires a more intricate integration strategy, as we detail in the following.

To render the dispersion integral over Eq.~\eqref{G_spectral_integral} well defined, we need to be able to interpret integrals of the form
\begin{align}
    I_\pm(s) &= \int_{\sth}^{\infty} ds^\prime \, \frac{T(s^\prime)}{s^\prime - s \mp i  \eps} \Big[ c_1 d_1(s^\prime) + c_2 d_2(s^\prime) \Big], \notag\\
    d_1(s) &= \frac{\log \big|\frac{s}{\sps}-1\big| - i \pi \theta(\sps-s)}{s-\sps},\qquad d_2(s) = \frac{1}{s-\sps},
\end{align}
where $T(s)$ is a regular function along the cut and for the pseudothreshold $\sps \mapsto \sps + i  \delta$ is implied, whenever necessary (see Refs.~\cite{Colangelo:2025ivq,Colangelo:2025iad} for similar integrals). To treat both the singularity stemming from the Cauchy kernel $1/(s^\prime-s\mp i \eps)$ as well as those introduced by the functions $d_{1,2}(s)$, we proceed similarly to Ref.~\cite{Stamen:2022eda} and distinguish between the following cases,
\begin{enumerate}
    \item $s<\sth$ or $\Im s\neq 0$,
    \item $\sth<s<\sps$,
    \item $s>\sps$.
\end{enumerate}

\begin{figure*}[t]
	\centering
	\includegraphics[width=0.35\linewidth]{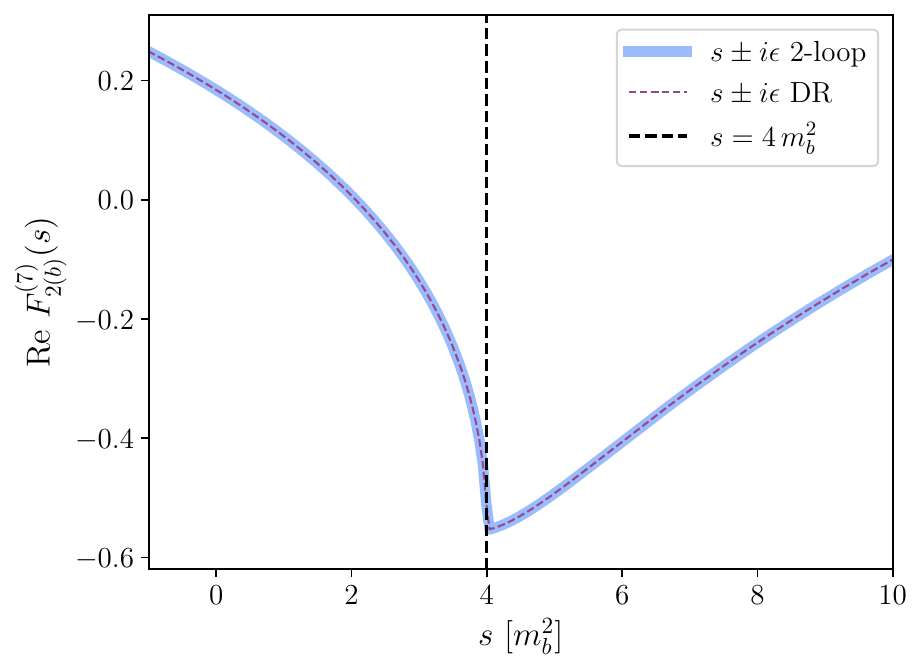}\quad
	\includegraphics[width=0.35\linewidth]{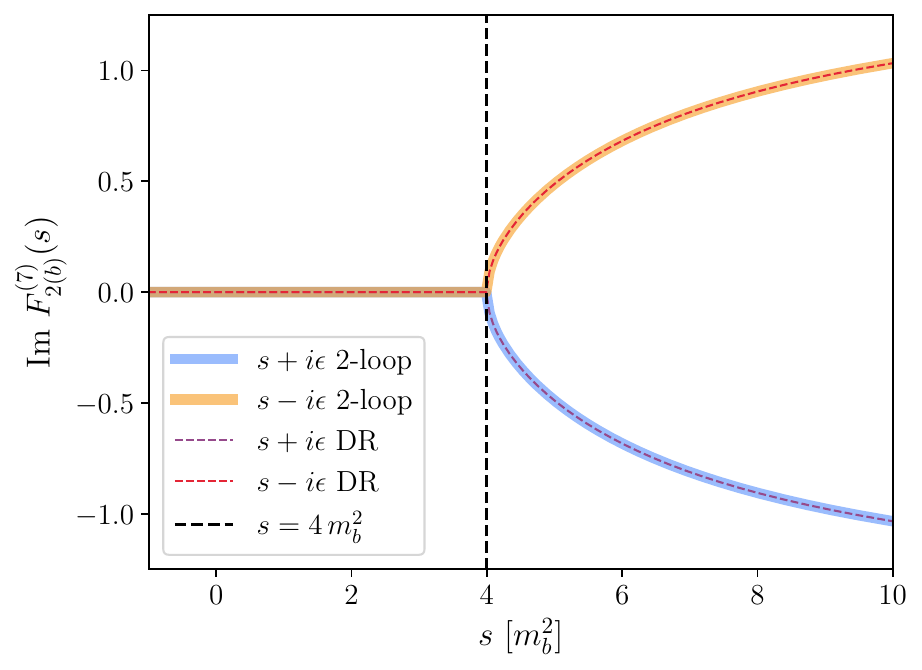}\\
	\includegraphics[width=0.35\linewidth]{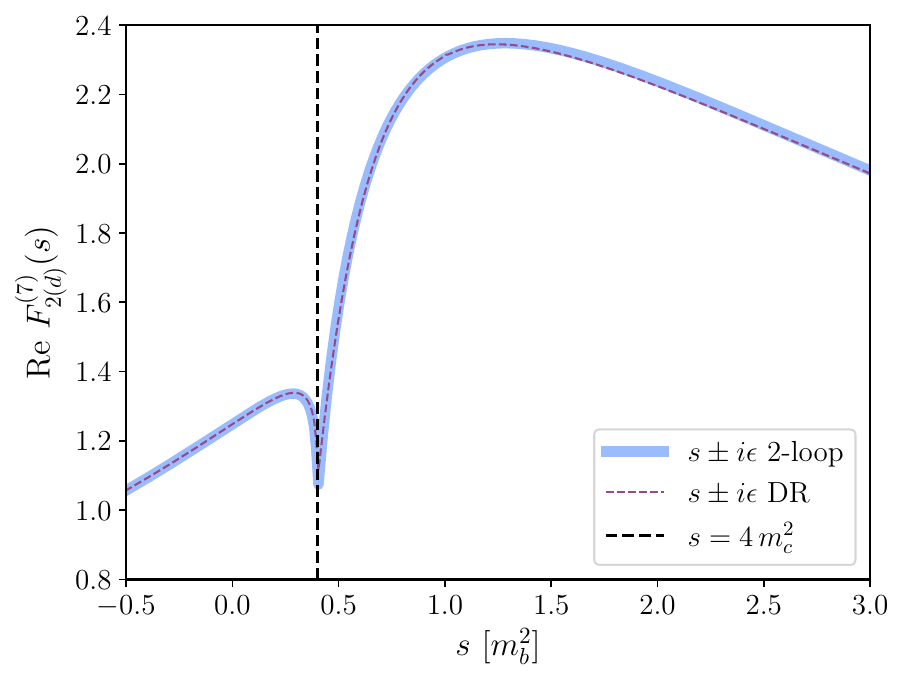}\quad
	\includegraphics[width=0.35\linewidth]{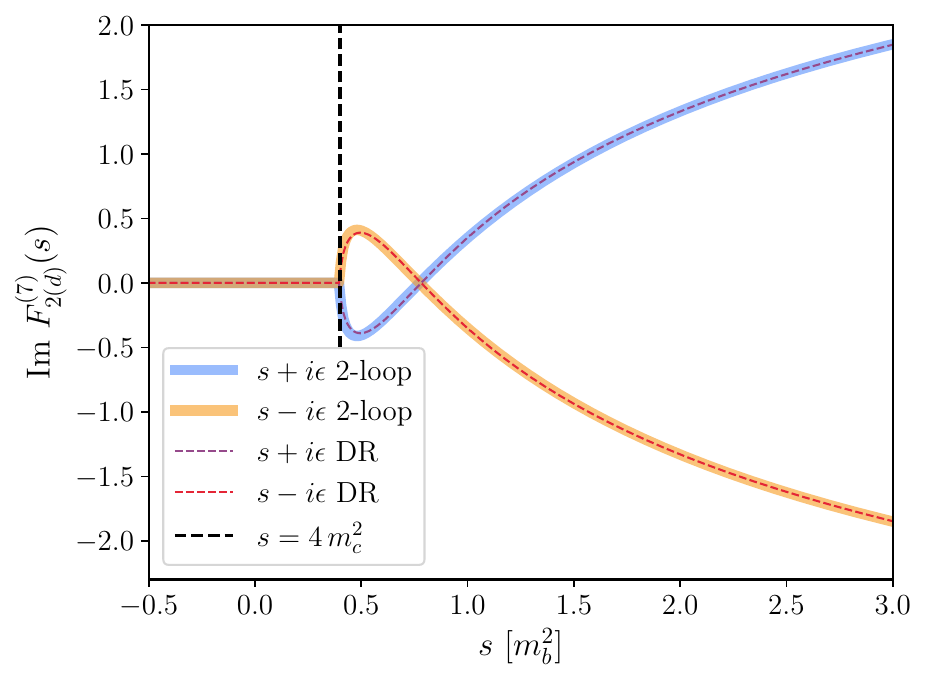}\\
	\caption{Real and imaginary part of the FFs for diagrams $(b)$ and $(d)$, comparing our results from the dispersion relation (``DR'') to the exact results from Ref.~\cite{Asatrian:2019kbk} (``$2$-loop''), in analogy to Fig.~\ref{fig:dispersion_relation_ac}. }
	\label{fig:dispersion_relation_bd}
\end{figure*}

\begin{figure*}[t]
	\centering
	\includegraphics[width=0.35\linewidth]{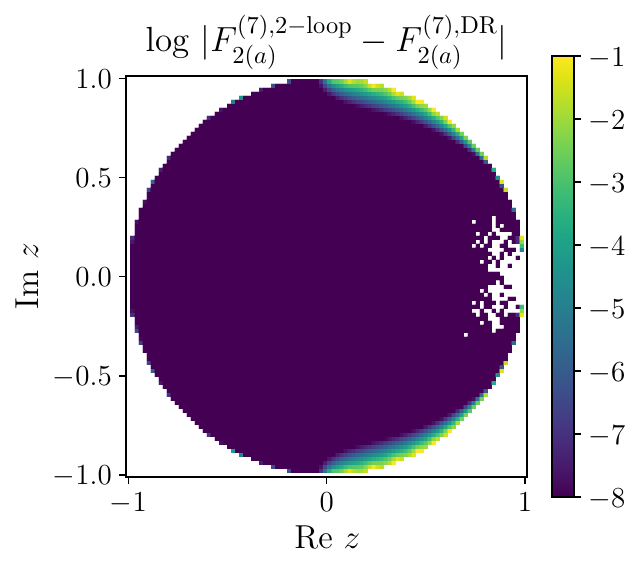}\quad
	\includegraphics[width=0.35\linewidth]{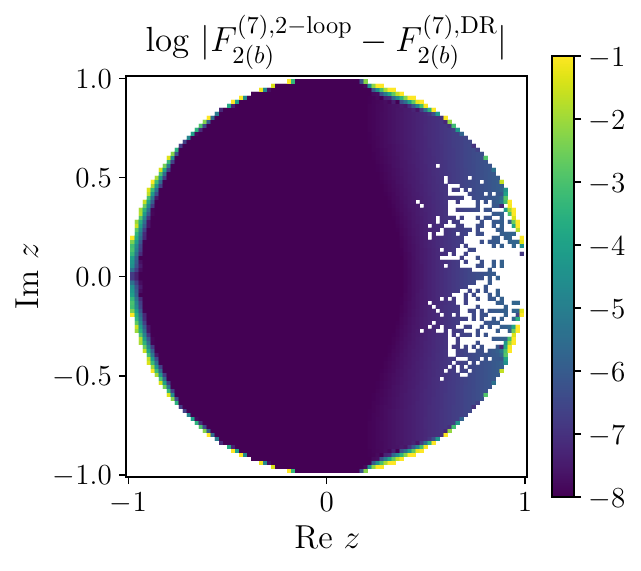}\\
	\includegraphics[width=0.35\linewidth]{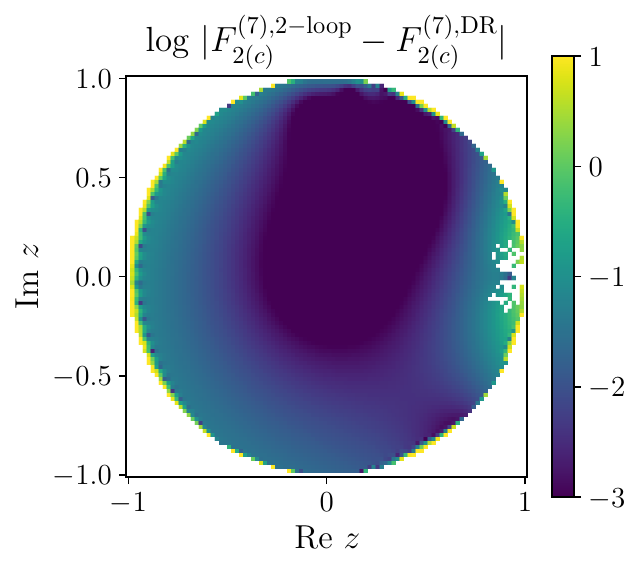}\quad
	\includegraphics[width=0.35\linewidth]{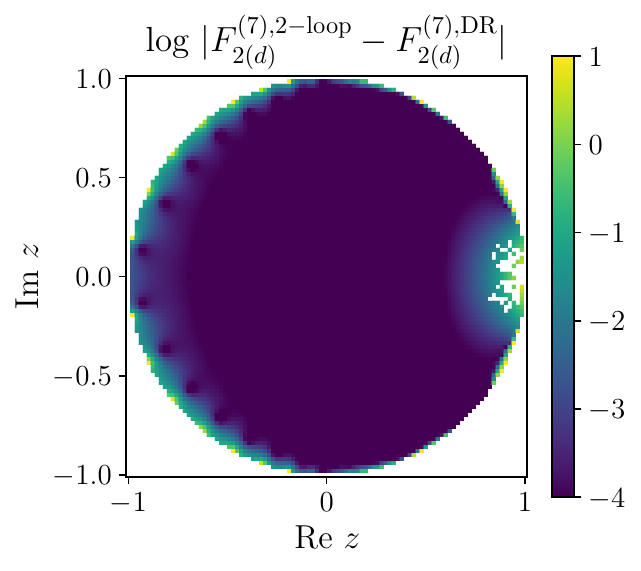}\\
	\caption{Deviations between our results from the dispersion relation (``DR'') and the exact results from Ref.~\cite{Asatrian:2019kbk} (``$2$-loop''), on a logarithmic scale and after conformal mapping $z(s)$ onto the unit circle (using the respective value of $\sth$ and $s_0=0$, except for $s_0=-m_c^2$ in the case of diagram $(a)$). For large values of $s$, instabilities occur in the implementation in the \texttt{EOS} software~\cite{EOSAuthors:2021xpv}, version 1.0.16~\cite{danny_van_dyk_2025_15783309}, corresponding to the white points in the figures.}
	\label{fig:dispersion_relation_complex}
\end{figure*}

Further, we introduce the following auxiliary integrals
\begin{align}
    Q_{1,2}(s,x,y) &= \int_{x}^{y} d s^\prime \, \frac{d_{1,2}(s^\prime)}{s^\prime-s},  & \!\!\text{for } & \sps \in [x,y] \text{ and } s \notin [x,y],\notag\\
    R_{1,2}^{\pm}(s,x,y) &= \int_{x}^{y} d s^\prime \frac{d_{1,2}(s^\prime)}{s^\prime-s\mp i \eps},  & \!\!\text{for } & \sps < x \text{ and } s \in [x,y],\notag\\
    S_{1,2}^{\pm}(s,x,y) &= \int_{x}^{y} d s^\prime \frac{d_{1,2}(s^\prime)}{s^\prime-s\mp i \eps},  & \!\!\text{for } & \sps > y \text{ and } s \in [x,y],
\end{align}
which yield
\begin{align}
    Q_1(s,x,y) &= \frac{1}{2(s-\sps)} \Bigg\{ 2 \biggl[ \dilog{\frac{y-\sps}{s-\sps}} - \dilog{\frac{x-\sps}{s-\sps}} \biggr]\notag\\
    &- \log\bigg(\frac{y}{\sps}-1\bigg) \biggl[ \log\bigg(\frac{y}{\sps}-1\bigg) - 2 \log\bigg(\frac{y-s}{\sps-s}\bigg) \biggr] \notag\\
    &+ \biggl[ \log\bigg(1-\frac{x}{\sps}\bigg) - i \pi \biggr] \notag\\
    &\quad\times\biggl[ \log\bigg(1-\frac{x}{\sps}\bigg) - 2 \log\bigg(\frac{x-s}{\sps-s}\bigg) - i \pi \biggr] \Bigg\}, \notag\\
    R_1^\pm(s,x,y) &= \frac{1}{2(s-\sps)} \Bigg\{ -2 \biggl[ \dilog{\frac{s-\sps}{y-\sps}} + \dilog{\frac{x-\sps}{s-\sps}}\notag\\
    &\quad+ \frac{1}{2} \biggl( \log\bigg(\frac{y-\sps}{s-\sps}\bigg) \pm  i \pi \biggr)^2 + \frac{\pi^2}{6} \biggr] \notag\\
    &- \log\bigg(\frac{y}{\sps}-1\bigg) \biggl[ \log\bigg(\frac{y}{\sps}-1\bigg) - 2 \log\bigg(\frac{y-s}{s-\sps}\bigg) \mp 2 i \pi \biggr]\notag\\
    &+ \log\bigg(\frac{x}{\sps}-1\bigg) \biggl[ \log\bigg(\frac{x}{\sps}-1\bigg) - 2 \log\bigg(\frac{s-x}{s-\sps}\bigg) \biggr] \Bigg\}, \notag\\
    S_1^\pm(s,x,y) &= \frac{1}{2(s-\sps)} \Bigg\{ 2 \biggl[ \dilog{\frac{\sps-y}{\sps-s}} + \dilog{\frac{\sps-s}{\sps-x}} \notag\\
    &\quad+ \frac{1}{2} \biggl( \log\bigg(\frac{\sps-x}{\sps-s}\bigg) \mp  i \pi \biggr)^2 + \frac{\pi^2}{6} \biggr] \notag\\
    &- \biggl[ \log\bigg(1-\frac{y}{\sps}\bigg) - i \pi \biggr] \notag\\
    &\qquad\times\biggl[ \log\bigg(1-\frac{y}{\sps}\bigg) - 2 \log\bigg(\frac{y-s}{\sps-s}\bigg) - i \pi \biggr] \notag\\
    &+ \biggl[ \log\bigg(1-\frac{x}{\sps}\bigg) - i \pi \biggr]\notag\\ &\quad\times\biggl[ \log\bigg(1-\frac{x}{\sps}\bigg) - 2 \log\bigg(\frac{s-x}{\sps-s}\bigg) - (1 \mp 2) i \pi \biggr] \Bigg\}, \notag\\
    Q_2(s,x,y) &= \frac{1}{s-\sps} \Biggl[ \log\bigg(\frac{y-s}{x-s}\bigg) - \log\bigg(\frac{y-\sps}{\sps-x}\bigg) - i \pi \Biggr], \notag\\
    R_2^\pm(s,x,y) &= S_2^\pm(s,x,y) \notag\\
    &= \frac{1}{s-\sps} \Biggl[ \log\bigg(\frac{y-s}{s-x}\bigg) - \log\bigg(\frac{\sps-y}{\sps-x}\bigg) \pm i \pi \Biggr].
\end{align}
In case 1 we only need to treat the pseudothreshold singularity and upon introducing a finite high-energy cutoff $\Lambda^2$ we find
\begin{align}
    I_\pm(s) &= \int_{\sth}^{\Lambda^2} d s^\prime \, \frac{T(s^\prime) - T(\sps)}{s^\prime-s} \Big[ c_1 d_1(s^\prime) + c_2 d_2(s^\prime) \Big] \notag\\
    &+ T(\sps) \Bigl[ c_1 \, Q_1\bigl(s,\sth,\Lambda^2\bigr) + c_2 \, Q_2\bigl(s,\sth,\Lambda^2\bigr) \Bigr].
\end{align}
In case 2 both the pseudothreshold singularity and the Cauchy kernel singularity need to be treated and we therefore split the integral into two parts by introducing the splitting point $p(s) = (s + \sps)/2$, to obtain
\begin{align}
    I_\pm(s) &= \int_{\sth}^{p(s)} d s^\prime \, \frac{T(s^\prime) - T(s)}{s^\prime-s} \Big[ c_1 d_1(s^\prime) + c_2 d_2(s^\prime) \Big]\notag\\
    &\quad+ T(s) \Bigl[ c_1 \, S_1^\pm\bigl(s,\sth,p(s)\bigr) + c_2 \, S_2^\pm\bigl(s,\sth,p(s)\bigr) \Bigr] \notag\\
    &+\int_{p(s)}^{\Lambda^2} d s^\prime \, \frac{T(s^\prime) - T(\sps)}{s^\prime-s} \Big[ c_1 d_1(s^\prime) + c_2 d_2(s^\prime) \Big] \notag\\
    &\quad+ T(\sps) \Bigl[ c_1 \, Q_1\bigl(s,p(s),\Lambda^2\bigr) + c_2 \, Q_2\bigl(s,p(s),\Lambda^2\bigr) \Bigr].
\end{align}
Similarly, in case 3 we find
\begin{align}
    I_\pm(s) &= \int_{\sth}^{p(s)} d s^\prime \, \frac{T(s^\prime) - T(\sps)}{s^\prime-s} \Big[ c_1 d_1(s^\prime) + c_2 d_2(s^\prime) \Big] \notag\\
    &\quad+ T(\sps) \Bigl[ c_1 \, Q_1\bigl(s,\sth,p(s)\bigr) + c_2 \, Q_2\bigl(s,\sth,p(s)\bigr) \Bigr] \notag\\
    &+\int_{p(s)}^{\Lambda^2} d s^\prime \, \frac{T(s^\prime) - T(s)}{s^\prime-s} \Big[ c_1 d_1(s^\prime) + c_2 d_2(s^\prime) \Big] \notag\\
    &\quad+ T(s) \Bigl[ c_1 \, R_1^\pm\bigl(s,p(s),\Lambda^2\bigr) + c_2 \, R_2^\pm\bigl(s,p(s),\Lambda^2\bigr) \Bigr].
\end{align}

\section{Numerical results of the dispersion relations}
\label{app:dispersion_relations_results}

Numerical results of the dispersion relation on the real axis were already provided for diagrams $(a)$ and $(c)$ in Fig.~\ref{fig:dispersion_relation_ac}. Figure~\ref{fig:dispersion_relation_bd} gives the analogous results for diagrams $(b)$ and $(d)$, in which cases the discontinuity is purely imaginary, so that for the real parts the limits $s\pm i\eps$ coincide. For the imaginary parts, a discontinuity only arises above the normal threshold.

We can also evaluate the dispersion relation in the complex plane away from the real axis, in which case the numerical accuracy tends to improve. This is illustrated in Fig.~\ref{fig:dispersion_relation_complex} for diagrams $(a)$--$(d)$, mapping the complex plane onto the unit circle via
\begin{equation}
 z(s)=\frac{\sqrt{\sth-s}-\sqrt{\sth-s_0}}{\sqrt{\sth-s}+\sqrt{\sth-s_0}}.
\end{equation}
While the numerical precision could be further improved by allowing for more terms in the parameterization of the discontinuities and finer integration grids, Figs.~\ref{fig:dispersion_relation_ac}, \ref{fig:dispersion_relation_bd}, and~\ref{fig:dispersion_relation_complex} demonstrate that the dispersion relations derived from the pertinent triangle topologies indeed hold, even in the presence of anomalous thresholds.

\bibliographystyle{apsrev4-1_mod}
\balance
\biboptions{sort&compress}
\bibliography{ref}

\end{document}